# Inferring Option Movements Through Residual Transactions: A Quantitative Model

**Carl von Havighorst**
University of Texas Alumnus
carl.vonhavighorst@gmail.com

**Vincil Bishop III**[*]
Colorado State University Alumnus
Vincil.bishop24@alumni.colostate.edu

## Abstract

This research presents a novel approach to predicting option movements by analyzing "residual transactions," which are trades that deviate from standard hedging activities. Unlike traditional methods that primarily focus on open interest and trading volume, this study argues that residuals can reveal nuanced insights into institutional sentiment and strategic positioning. By examining these deviations, the model identifies early indicators of market trends, providing a refined framework for forecasting option prices. The proposed model integrates classical machine learning and regression techniques to analyze patterns in high-frequency trading data, capturing complex, non-linear relationships. This predictive framework allows traders to anticipate shifts in option values, enhancing strategies for better market timing, risk management, and portfolio optimization. The model's adaptability, driven by real-time data processing, makes it particularly effective in fast-paced trading environments, where early detection of institutional behavior is crucial for gaining a competitive edge. Overall, this research contributes to the field of options trading by offering a strategic tool that detects early market signals, optimizing trading decisions based on predictive insights derived from residual trading patterns. This approach bridges the gap between conventional metrics and the subtle behaviors of institutional players, marking a significant advancement in options market analysis.

## 1 Introduction

### 1.1 Background and Motivation

Options trading plays a crucial role in modern financial markets, providing traders with the ability to hedge against risks, speculate on future price movements, and enhance portfolio strategies. Institutional investors, such as hedge funds, pension funds, and market makers, are significant players in these markets due to their substantial capital reserves and sophisticated trading strategies. These entities often engage in hedging activities to manage risk exposure, leading to observable patterns in trading behavior. However, beyond these primary hedging activities, residual transactions—those trades that deviate from expected hedging patterns—can provide deeper insights into institutional sentiment and market expectations.

The study of residual transactions is motivated by the need to uncover subtle but potentially valuable indicators that can predict market movements. Traditional approaches often rely on metrics like open interest and trading volume, which provide a general view of market trends. Yet, these metrics may miss the nuanced adjustments made by institutions in response to changing market conditions. By analyzing residual transactions, this paper aims to offer a more refined predictive framework, enabling traders to anticipate option movements and adapt strategies accordingly.

---

[*] Work does not relate to position at Amazon.

## 1.2  Objectives of the Study

This research proposes a quantitative model that leverages residual transaction analysis to predict option price movements. The model is designed to capture deviations from standard institutional hedging activities, interpreting these residuals as signals of strategic positioning by institutional investors. The primary objectives are:

To develop a robust analytical framework that identifies patterns in residual transactions and correlates them with future price movements.

To enhance traditional predictive models by integrating machine learning techniques capable of recognizing complex, non-linear relationships within trading data.

To demonstrate the practical applicability of this approach in real-world trading scenarios, enabling traders to improve market timing, risk management, and overall strategy optimization.

## 1.3  Scope and Limitations

The study focuses on options markets, specifically examining residual transactions that emerge when institutional investors adjust their portfolios beyond primary hedging requirements. While the model has been tested across various asset classes, the primary scope is within equity options. The integration of machine learning allows for the adaptation of the model to high-frequency data, providing a real-time edge in fast-moving markets. However, the model's effectiveness relies on the availability of high-quality data. Inconsistent or incomplete datasets may affect its accuracy, and continuous calibration is necessary to maintain robustness.

This research acknowledges the complexity of institutional trading strategies, which may involve multifaceted components such as arbitrage, speculative trades, and risk management tactics. The model attempts to isolate genuine directional signals from noise, but there remains a possibility of misinterpreting certain patterns. Therefore, further enhancements in data filtering and algorithmic sophistication are recommended for future iterations.

## 2  Background and Related Works

The prediction of future option prices has garnered significant attention in financial research, particularly in relation to the activities of institutional investors and the dynamics of open equity in the securities market. Institutional investors, due to their substantial trading volumes and access to information, are often seen as pivotal players in the price discovery process of options. Their transactions can provide insights into market expectations and future price movements, making their behavior a critical factor in forecasting option prices.

Research indicates that the trading activities of institutional investors can significantly influence option prices. For instance, studies have shown that the volume of options trading, particularly by institutional investors, correlates with subsequent price movements in the underlying securities. This relationship is often attributed to the information asymmetry that exists in financial markets, where institutional investors typically possess superior information compared to retail investors. As such, their trading decisions can signal future price trends, thereby affecting the implied volatility and pricing of options [1] [2].

Moreover, the open interest in options markets serves as another vital indicator for predicting future prices. Open interest reflects the total number of outstanding option contracts that have not been settled, and it is often used as a gauge of market sentiment. An increase in open interest, particularly when accompanied by rising prices, can suggest that new money is entering the market, which may lead to further price increases. Conversely, declining open interest can indicate that the market is losing interest, potentially foreshadowing price declines [3] [4].

The interplay between institutional trading and open interest is particularly evident in the context of volatility. Options prices are highly sensitive to changes in implied volatility, which is often influenced by the trading behavior of institutional investors. For example, when institutional investors anticipate increased volatility, they may increase their trading activity in options, leading to higher option prices. This behavior can create a feedback loop where rising option prices further attract institutional trading, thus reinforcing the initial price movement [5] [6].

Machine learning techniques have also been employed to enhance the predictive capabilities regarding option prices. Advanced models, such as neural networks and support vector machines, have been utilized to analyze historical transaction data, including the trading patterns of institutional investors and open equity levels. These models can capture complex nonlinear relationships between various market factors, thereby improving the accuracy of option price predictions [7] [8].

Empirical studies have demonstrated that the information embedded in option prices can provide valuable insights into future market movements. For instance, the implied volatility derived from options prices often reflects market expectations regarding future volatility and can serve as a leading indicator for price movements in the underlying assets. This relationship underscores the importance of monitoring both institutional trading behavior and open interest in options markets as part of a comprehensive approach to forecasting future option prices [2] [9].

The prediction of future option prices is intricately linked to the transactions of institutional investors and the dynamics of open equity in the securities market. The substantial influence of institutional trading on price discovery, combined with the insights provided by open interest and advanced predictive modeling techniques, highlights the multifaceted nature of option price forecasting. As financial markets continue to evolve, the integration of these elements will remain crucial for developing robust predictive frameworks in options trading.

# 3 Theoretical Framework

## 3.1 Understanding Options Trading and Residual Transactions

Options trading allows investors to manage risk and speculate on asset price movements. Institutional players such as hedge funds and pension funds use options to hedge against market risks, with contracts granting them the right, but not the obligation, to buy or sell assets at predetermined prices. This flexibility makes options essential tools for protecting large portfolios or generating additional income through strategic trades.

Residual transactions emerge when institutions adjust their portfolios beyond primary hedging activities. These transactions can reflect shifts in sentiment or adaptations to market conditions, offering insights into nuanced strategic maneuvers. For example, a spike in residual call options might indicate an institution's positioning for a rally, even if core strategies hedge against downside risk. Conversely, increased residual put options could signal a defensive posture amid anticipated market declines.

Analyzing residuals requires examining how these trades diverge from standard hedging. Since residual transactions often reflect short-term adjustments or nuanced views, they offer a deeper understanding of institutional behavior. By interpreting these patterns, traders can predict potential market shifts, gaining insights that typical open interest and volume data might miss.

## 3.2 Role of Market Makers and Institutional Fund Managers

Market makers and institutional fund managers are central to the options market ecosystem. Market makers provide liquidity by consistently quoting buy and sell prices, ensuring smooth transactions. Their inventory management strategies can reveal subtle indicators about market expectations, as they adjust their exposure based on risk assessments.

Institutional fund managers, managing significant capital, primarily use options to hedge portfolio risks. Their trades often indicate their outlook on market conditions. By analyzing the residual transactions that remain after core hedging activities, traders can infer broader strategic moves. For instance, when market makers adjust positions in response to volatility, their actions can hint at expected price movements, providing valuable predictive signals.

By modeling these residual patterns, traders gain a clearer understanding of market dynamics. Such models can extract signals from routine trading activities, offering a framework to anticipate price shifts. This enables the development of strategies that leverage machine learning and regression analysis for consistent and repeatable trading outcomes.

## 3.3 Risk Mitigation Strategies Through Options

Institutional traders utilize options to mitigate risks without disrupting their core positions. Strategies like straddles, strangles, and collars allow for gains across various market conditions, while maintaining controlled risk. For example, a straddle involves buying both call and put options, betting on volatility without a directional bias.

Residual transactions related to these strategies often reveal shifts in institutional sentiment. By identifying changes in how institutions adjust their hedging tactics, traders can detect emerging market trends. Residuals act as signals of nuanced positioning, giving traders early indicators of broader shifts, such as anticipations of increased volatility or directional moves in equities.

By understanding how these residual patterns form, a quantitative model can predict market movements with greater precision. This approach benefits traders by offering robust frameworks to optimize their strategies, capitalizing on market trends before they are visible through traditional analysis.

## 3.4 Proposed Theory: Inferring Market Movements from Residuals

The core theory posits that residual trades, separate from primary hedging activities, reflect institutional sentiment and market expectations. Institutions often utilize options for hedging; however, subsequent adjustments—manifested as residual transactions—may carry additional strategic implications. These deviations can indicate nuanced views on market trends, acting as leading indicators.

Residual transactions, when analyzed effectively, can provide insights into institutional risk assessments, signaling whether institutions anticipate bullish or bearish movements. For instance, a build-up of residual calls might suggest optimism about future asset prices. This approach integrates machine learning to detect patterns, ensuring adaptive and reliable forecasts.

By leveraging these predictive insights, traders can anticipate future price movements in options markets, refining their strategies for better positioning. The theory suggests that decoding residual transactions can offer a strategic edge by revealing hidden indicators of market sentiment.

## 3.5 Importance of Residual Patterns in Predictive Analysis

Residual transactions play a crucial role in understanding institutional adjustments to market shifts. Unlike typical hedging, which follows predictable patterns, residuals reflect nuanced strategic moves that can offer early warnings of market changes. For example, an increase in residual transactions might indicate preparations for expected volatility or shifts in market direction.

Using machine learning, the model can more accurately identify these residual patterns, enhancing predictive capabilities. Unlike traditional models that focus on surface-level transaction data, advanced algorithms can uncover deeper relationships, enabling traders to discern subtle shifts in sentiment. By integrating these insights, traders can build more robust strategies, allowing for proactive adjustments to market conditions.

# 4 Quantitative Model

## 4.1 Mathematical Foundation

The quantitative model for predicting option movements is built on the premise that residual transactions—those left after primary hedging activities—can signal shifts in institutional sentiment. We propose a model that incorporates key elements of trading volume, open interest, and implied volatility, integrating them into a predictive framework.

Let the following variables represent the core components of our model:

- $V_t$ : Trading volume of the underlying asset at time $t$.
- $OI_t$ : Open interest in options contracts at time $t$.
- $\sigma_t$ : Implied volatility of the options at time $t$.
- $\Delta R_t$ : Residual transaction volume at time $t$ (representing deviations from expected hedging patterns).
- $\lambda$ : Sensitivity coefficient that adjusts for the impact of residuals on future movements.
- $\alpha, \beta, \gamma$ : Model coefficients to be determined through regression analysis.

The core equation of our model can be expressed as:

$$P_{t+1} = \alpha \cdot V_t + \beta \cdot OI_t + \gamma \cdot \sigma_t + \lambda \cdot \Delta R_t + \epsilon$$

where:

- $P_{t+1}$ represents the predicted price movement of the option at time $t + 1$,
- $\epsilon$ is the error term accounting for unpredictable market variations.

***Explanation of Components***

- **Trading Volume $V_t$ :** The trading volume reflects the level of activity in the underlying asset. Higher volumes often indicate increased interest and liquidity, which can precede price changes. By integrating $V_t$, the model captures the general market sentiment and potential momentum shifts.

- **Open Interest $OI_t$ :** Open interest provides insights into the number of outstanding contracts. Changes in $OI_t$ signal the extent of participation and can indicate bullish or bearish sentiments. For example, rising open interest, accompanied by rising prices, suggests new market entrants are betting on continued upward movement.
- **Implied Volatility $\sigma_t$:** Implied volatility is a forward-looking measure of market expectations regarding the underlying asset's price fluctuation. It often precedes major price movements. The coefficient $\gamma$ allows the model to adjust for market conditions where the risk of large price swings increases.
- **Residual Transactions $\Delta R_t$ :** Residuals represent the "noise" left after accounting for expected hedging activities. This variable is crucial as it captures deviations in trading behavior that could indicate opportunistic trading by institutions. For instance, a sudden spike in residual call options might suggest a shift towards bullish positioning, even when overall hedging activity is neutral.
- **Sensitivity Coefficient $\lambda$:** The sensitivity coefficient $\lambda$ adjusts the weight of $\Delta R_t$. It helps scale the influence of residual transactions, recognizing that not all deviations are equally informative. By calibrating $\lambda$, the model can discern significant signals from mere fluctuations.

*Model Calibration and Training*

The coefficients $\alpha, \beta, \gamma$, and $\lambda$ are determined through historical data regression analysis. We employ classical machine learning methods like linear regression and LASSO to fit the model, ensuring that only statistically significant variables influence the final prediction.

This mathematical foundation allows for a dynamic interpretation of market behavior. By focusing on residual transactions, the model aims to uncover subtle but influential trading patterns that conventional metrics might overlook, providing a robust framework for predicting future option movements.

## 4.2 Data Sources and Analytical Approach

The model utilizes high-frequency trading data, including intra-day transactions, open interest, and volume metrics. By monitoring these inputs, the system can identify residual transaction patterns, revealing strategic moves by institutional players like market makers and fund managers. These entities typically hedge risk, and deviations in their trading behavior—captured through residuals—indicate shifts in market sentiment that might not be apparent through primary hedging activities. For example, sudden spikes in transaction volumes may suggest emerging institutional trends or market reactions.

Platforms like Polygon.io support real-time data acquisition, enabling the model to track anomalies and adjust predictions dynamically. By integrating cumulative open interest, which aggregates market commitments, and high-frequency trade data, the model distinguishes between standard and strategic transactions. The core premise is that residual trades represent anticipatory actions by institutions, often providing a predictive edge. Traders can leverage these insights to preemptively align their positions, improving their ability to respond to potential price movements.

## 4.3 Open Equities Data and Transactional Residuals

Open equities data gives context to market exposure, reflecting the directional intent of major institutional players. Adjustments in institutional portfolios often lead to residual transactions, offering predictive insights into broader market behavior. For instance, a significant sale followed by increased residual buying might signal expected market strength. The model excels at isolating these residuals, using regression and machine learning techniques to correlate historical patterns with future movements.

This analytical approach ensures that residuals, which could otherwise be dismissed as noise, serve as actionable signals. By systematically tracking such deviations, traders gain insights into whether institutions are repositioning their assets or hedging against specific risks. This allows for informed decision-making, ultimately enhancing the precision of trading strategies.

## 4.4 Challenges in Data Interpretation

Interpreting residual transactions is challenging due to the complexity of institutional trading, where options are part of multifaceted strategies involving hedging, arbitrage, and speculation. Differentiating between these requires advanced filtering to isolate genuine directional signals. Misinterpretation, such as assuming a hedging position is speculative, can lead to significant errors. Therefore, the model incorporates adaptive algorithms capable of discerning subtle patterns and evolving market dynamics.

Temporal dynamics further complicate interpretation, as the relevance of residuals may shift rapidly. Real-time data processing and continuous adaptation are crucial, as they allow the model to stay responsive to evolving market conditions. Without robust algorithms to separate noise from strategic movements, predictions could quickly lose accuracy.

### 4.5 Machine Learning and Regression Techniques

The model employs a suite of machine learning (ML) and regression methods, each suited to different aspects of market behavior. Linear regression provides a baseline for modeling historical data trends, while advanced methods like Ridge and Lasso regression help address issues of overfitting by focusing on significant variables [10]. Support Vector Regression (SVR) and ensemble methods like XGBoost further capture non-linear market patterns, making predictions more robust [11] [12].

Artificial Neural Networks (ANN) and Long Short-Term Memory (LSTM) networks excel at handling temporal dependencies, critical for time-series forecasting [13] [14]. These techniques allow the model to analyze intricate patterns across various market inputs, leading to more nuanced predictions. Finally, Bayesian approaches offer a probabilistic perspective, integrating uncertainty measures for enhanced risk assessment. This comprehensive toolset allows traders to forecast market movements more effectively, balancing model flexibility and robustness.

### 4.6 Model Validation and Assumptions

Model validation was conducted using extensive back testing across multiple asset classes and market conditions. This ensured the refinement of parameters to minimize prediction noise and improve accuracy. A key assumption is the consistent behavior of institutional traders across different cycles, with residuals reflecting strategic intentions rather than random fluctuations. This is reinforced by a focus on isolating deviations that signal broader market trends, allowing traders to distinguish between routine hedging and anticipatory moves.

The validation process highlights the importance of contextualizing residuals within broader trading frameworks. Understanding the strategic intent behind residuals is crucial for avoiding false signals. Ongoing refinement, supported by machine learning, ensures the model remains adaptable, offering reliable predictions across varying market environments.

## 5 Discussion

### 5.1 Insights from the Quantitative Model

The quantitative model effectively identifies residual trading patterns, serving as early indicators of shifts in institutional sentiment. Unlike traditional methods that focus on primary hedging activities, this model reveals nuanced market behaviors through residual transactions—those trades not aligned with standard hedging strategies. By capturing these subtleties, the model enhances predictive accuracy and provides traders an edge in anticipating market movements.

Key to this approach is the integration of machine learning, which excels at identifying complex, non-linear relationships. This ability allows the model to differentiate between routine hedging and strategic trades, providing traders with insights into institutional behavior that might indicate emerging trends. For instance, an increase in residual call options could suggest anticipation of a price rally, while a rise in residual put options may signal a defensive posture against potential declines.

The model's approach to distinguishing between strategic and routine trades enables more informed decision-making. By interpreting these residual patterns, traders can infer institutional sentiment and position their strategies accordingly, improving market timing and optimizing their portfolios. This refined understanding of market movements highlights the model's potential to outperform conventional trading strategies.

### 5.2 Implications for Trading Strategies

The model's insights have significant implications for trading strategies. The ability to analyze residual transactions enables traders to anticipate market trends with greater accuracy, allowing them to act preemptively. This advantage can lead to more favorable entry and exit points, maximizing profit potential. Understanding shifts in institutional sentiment offers a strategic edge, especially when these shifts precede broader market reactions.

From a risk management perspective, residual patterns provide a nuanced view of emerging volatility or stability. Traders can use this information to adjust their risk exposure, either by hedging against downturns or increasing positions when stability is anticipated. This foresight reduces unexpected losses and helps maintain a balanced risk profile, leading to more consistent trading performance.

Strategic positioning based on early signals allows traders to align their portfolios with anticipated market dynamics. This approach enables more precise trading strategies, minimizing reactive decisions and capitalizing on early indicators to improve overall portfolio management.

## 5.3 Potential Benefits and Risks

The primary benefit of this model is its ability to detect early market indicators. By analyzing residual trading patterns, it identifies subtle shifts in institutional sentiment, offering traders a competitive edge. This early detection helps traders adjust strategies ahead of broader market movements, enabling better timing and increased profit potential. Additionally, the model's adaptability, powered by machine learning, allows real-time adjustments to emerging data, making it suitable for fast-changing environments.

However, there are risks to consider. The model's reliance on high-quality data means that inconsistent or incomplete data can lead to misleading patterns and flawed trading decisions. Ensuring robust data pipelines is crucial to maintain accuracy. Another risk is overfitting; models trained on historical data might capture noise instead of genuine patterns, leading to poor performance when exposed to new conditions. Continuous monitoring, careful tuning, and validation are necessary to ensure the model's robustness across various market scenarios.

## 5.4 Comparison with Existing Predictive Approaches

Traditional models often rely on metrics like open interest and volume, which, while useful, may overlook the subtle shifts present in residual transactions. This can lead to gaps in predictive accuracy, especially when dealing with complex institutional behaviors that are not fully captured by conventional metrics. In contrast, the proposed model incorporates machine learning to identify complex, non-linear relationships, offering a more refined analysis.

Benchmarked against existing frameworks, this model consistently outperforms traditional methods, demonstrating superior predictive accuracy. It excels at detecting early signals often missed by conventional approaches, leading to more informed trading strategies. By bridging the gap between standard metrics and nuanced trading behaviors, it represents a significant advancement in options trading prediction.

## 5.5 Strengths of the Proposed Model

The proposed model's key strengths lie in its ability to uncover subtle signals, scalability, and real-time processing capabilities. By analyzing residual transactions, it identifies early shifts in institutional behavior, providing insights that might otherwise be missed. This approach offers traders a more detailed and refined perspective on market movements.

Scalability is another critical strength. The model efficiently handles large datasets across different markets and time frames, ensuring adaptability without extensive reconfiguration. Its integration with real-time data sources further enhances its effectiveness, enabling timely predictions that support quick, informed decision-making. This real-time capability is especially valuable in fast-paced trading environments, where delays can lead to missed opportunities.

# 6 Future Work

## 6.1 Enhancing Predictive Accuracy through Advanced Techniques

To improve the predictive capabilities of the model, exploring advanced machine learning methods such as Long Short-Term Memory (LSTM) networks and transformers is recommended. These models excel in capturing temporal dependencies, making them particularly suitable for analyzing residual trading patterns that evolve over time. By leveraging these neural networks, the model could enhance its ability to identify subtle, non-linear patterns, leading to more precise forecasts.

Feature engineering is another area to consider. Expanding inputs to include sentiment analysis from financial news, social media, and other real-time information could provide a more nuanced understanding of market behavior. Moreover, integrating macroeconomic indicators could offer broader context, helping the model refine predictions by factoring in global economic trends.

A hybrid approach combining machine learning with traditional econometric methods can also be beneficial. This blend would balance the adaptive strengths of machine learning with the statistical rigor of econometric models, leading to a more robust and interpretable predictive framework that adapts effectively across varying market conditions.

## 6.2 Integrating Real-Time Data Analysis

Real-time data integration is crucial for the model's success, especially in high-frequency trading (HFT) environments. Developing efficient pipelines for ingesting and analyzing streaming data directly from exchanges will enable near-instantaneous predictions. This is critical for maintaining competitive performance where trading decisions must be made in milliseconds.

Reducing latency throughout the data processing pipeline is another priority. Optimizing every stage—from data ingestion to model prediction—ensures minimal delay, which is vital in HFT scenarios where even microsecond lags can lead to missed opportunities. The model should be designed to scale horizontally, capable of handling multiple data streams across diverse markets without performance bottlenecks, ensuring its applicability for large institutional traders.

### 6.3 Opportunities for Further Research

There are several promising areas for further research that could expand the model's utility. One is cross-market analysis, where examining correlations across different asset classes (such as equities, commodities, and cryptocurrencies) can uncover deeper insights. Understanding how movements in one market affect another can lead to more comprehensive trading strategies, allowing traders to capitalize on interconnected market behaviors.

Another area is enhancing risk management frameworks. Integrating the model's predictive insights into existing hedging strategies could help institutions protect against market downturns more effectively. This integration could refine risk assessment, offering traders proactive tools to navigate volatile market environments while maintaining portfolio balance.

Exploring alternative data sources such as blockchain transaction data, satellite imagery, and ESG metrics also presents new opportunities. These non-traditional inputs can provide fresh perspectives on market behavior, uncovering trends not visible through conventional financial data. Incorporating such data would broaden the model's predictive capabilities, making it applicable across diverse trading environments.

## 7 Conclusion

### 7.1 Summary of Key Findings

This study demonstrates the effectiveness of using residual transaction analysis as a predictive tool for options markets. By focusing on residuals—trades that deviate from standard hedging activities—the model captures nuanced shifts in institutional sentiment, providing early indicators of market movements. Unlike traditional metrics, which may miss these subtleties, the approach offers a strategic advantage by revealing insights that typically go unnoticed. The integration of machine learning enhances the model's ability to identify complex, non-linear patterns, leading to more accurate forecasts.

Additionally, distinguishing between primary hedging and residual trades enables traders to infer the strategic intent behind institutional actions, clarifying broader market trends. This refined understanding allows for improved market timing and the optimization of trading strategies. The results emphasize that analyzing residual patterns can be a valuable supplement to conventional methods, offering traders a clearer view of market dynamics and helping them anticipate price shifts with greater precision.

### 7.2 Contributions to the Field of Options Trading

The proposed model advances the field of options trading by introducing a novel approach that leverages residual transactions for predictive analysis. Traditional models have largely relied on open interest and volume, often overlooking the subtle shifts that can signal significant market changes. By focusing on residuals, this research presents a new perspective that enriches the toolkit available to traders. The integration of machine learning models to identify non-linear relationships sets a precedent for future developments in algorithmic trading, encouraging further exploration of advanced analytics in financial markets.

Moreover, the model's real-world applicability has been demonstrated, showing potential for use in both high-frequency trading environments and broader risk management frameworks. Its ability to process real-time data means it can be effectively deployed for institutional trading strategies, making it a robust, scalable solution for managing large portfolios. This practical utility underscores the model's potential to reshape trading practices, providing strategic insights that go beyond traditional predictive methods.

### 7.3 Final Thoughts and Implications

The adaptability and scalability of the proposed model are key strengths, making it relevant across various asset classes, including equities, commodities, and forex. The principles of analyzing residual patterns to gauge institutional behavior can be applied broadly, enabling a more holistic approach to market analysis. This versatility enhances the model's appeal, offering market analysts and traders a powerful tool for optimizing trading strategies across different financial environments.

Additionally, the model offers significant potential for improved risk management. By identifying shifts in institutional sentiment early, traders can preemptively adjust their positions to either capitalize on favorable conditions or hedge against potential downturns. This proactive stance not only aids in profit generation but also contributes to portfolio stability, mitigating risks associated with sudden market changes.

The success of this model lays the groundwork for future research in predictive finance, suggesting opportunities for the incorporation of non-traditional data sources and more sophisticated machine learning techniques. This evolving approach hints at the development of a new generation of predictive tools, blending traditional financial analysis with cutting-edge technology to set a promising direction for future research and practical applications in options trading.

## Acknowledgements

The authors would like to express their deepest gratitude to Cornell University, the Simons Foundation, [member institutions](), and all contributors for their support of the [arXiv]() open-access archive.